\documentclass[journal=jacsat,manuscript=article]{achemso}
\usepackage[version=3]{mhchem} 
\usepackage[T1]{fontenc}       
\usepackage{graphicx} 
\usepackage{subfig}
\usepackage{multirow}
\usepackage{longtable}
\usepackage{achemso} 
\usepackage{array,booktabs}
\usepackage{float}
\usepackage{caption}
\usepackage{color}
\usepackage{enumerate}
\usepackage{amsmath}
\usepackage[none]{hyphenat}

\author{Madhuri Mukhopadhyay}
\affiliation {Theoretical Sciences Unit, Jawaharlal Nehru Centre for Advanced Scientific Research, Bangalore 560064, India\\}
\author{Bradraj Pandey}
\affiliation {Theoretical Sciences Unit, Jawaharlal Nehru Centre for Advanced Scientific Research, Bangalore 560064, India\\}
\author{Swapan K Pati}
\affiliation {Theoretical Sciences Unit, Jawaharlal Nehru Centre for Advanced Scientific Research, Bangalore 560064, India\\}
\email{pati@jncasr.ac.in}

\title[An \textsf{achemso} demo]
 {Engineering the De-localized States of Graphene Quantum Dots}

\begin{document}

  \begin{figure} \centerline{{\bf {Graphical TOC entry}}}
\center\includegraphics[scale=1.2]{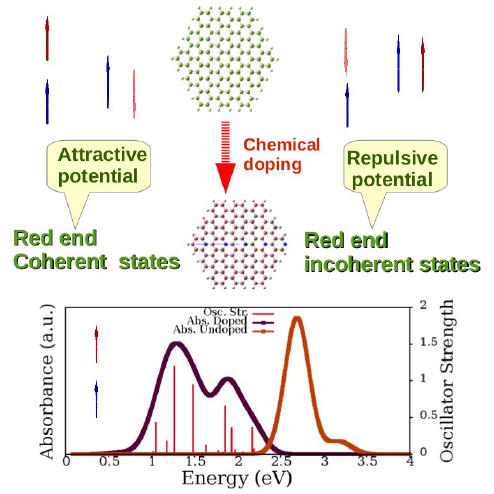}  \end{figure}


\begin{abstract}

Employing a combination of many-body configuration interaction method described by extended Hubbard model 
along with first principle calculations we predict  the emergence of high oscillator strength at near-IR region
which originates from the Davydov  type of splitting in doped graphene quantum dots (GQD). 
Incorporation of strain in GQD promotes closely spaced bright states inciting for coherent excitation.
 Controlling the destructive interference of the functionalized nano graphene quantum states, the dark states can be tuned 
towards red end ensuing  the system as a good candidate for photocell whereas coherent states can be tailored to 
concentrate the light at very high intensity resulting an opportunity for photonic device.
\end{abstract}

\section {SECTION:} Spectroscopy, Photochemistry, and Excited States

\newpage

\maketitle


\newpage

\section{INTRODUCTION}
With the  advancement in the fabrication of atomically thin materials \cite{r1,nat_tech_photodetect,r2,r3,r4}
and graphene nano flakes based devices, 
functionalized graphene  nano flakes are getting huge attention due to their promising electronic and optical 
properties, especially for applications in photonics and solar cells {\cite{natphoton2012,IEEE2013,natcom2015,r7}.
 The electronic and optical properties of the graphene based low dimensional 
structures change in non-trivial manner depending on the shape, size and edges \cite{r8, r9,many_jpc}.
 Stacking of the graphene layers controls the optical properties in significant ways. Depending on the angle of twisted sublayer of graphene it shifts the absorption 
spectral energy and gives van Hove singularities \cite{10prb205404}. Functionalization of graphene
 nano structures changes the properties in dramatic way in many respects \cite{r11,jpcl_gr_oxide}.
  For example, graphene has been predicted to be gap less semi-metal but fully hydrogenated graphene (graphane)
 has been predicted to have a LDA gap of 3.4eV \cite{12prb153401}. \\
 While graphene is stabilized due to delocalization, its B-N analogue is a stable Mott insulator due to strong interaction. 
However, since B-N is isoelectronic with C-C,
replacing C with B or N adds a  hole or electron, keeping the structure same.
 Hence, graphene has been doped with nitrogen and boron in several earlier
studies \cite{13prl036808,14prl196803}.
 However, reliable control of opto-electronic properties,  by chemical doping, is still a challenging task. Experimentally, one of 
the most common techniques for the production of doped graphene is the in situ introduction of B-(N-)containing precursors during a 
chemical vapor deposition (CVD) process \cite{15prb075401,16nanolet5401}. Other methods have also been used successfully to
produce nitrogen doped graphene fragments \cite{17prb161408,18scinece999}.  
There has been many works on the optical response of graphene  and  truncated graphene  $i.e.$ graphene quantum dots (GQD) and for GQD   
  the excitonic features are more exotic than their 2D analogue \cite{19prl186802,science2008_gqd}.
  Now the coherent control of excitonic states, over the optical transition is dictated by the 
  linear superposition of quantum states with symmetric or anti-symmetric phase matching resulting either addition or reduction  of 
  electrical transition dipole \cite{shukla,jpc_c}.\\
 The dipole dipole interaction  modulates the charge density wave and spin density wave phases, which can modify the 
 optical response in significant way. We have investigated here how the architecture of the boron nitrogen substitution in GQD can govern the Davydov type splitting \cite{21prl253601,acene_davy}
resulting in a coherent control of the excitonic states. The relative position of boron and nitrogen substitution and their partial dipole- partial dipole coupling 
can perturb the shallow bound exciton of GQD   and in some cases can overwhelm the intrinsic excitonic interaction and can develop 
 high oscillator strength at different energy making the  system interesting for optoelectronic devices. 
On the other hand, particular orientation of partial dipole created by boron and/or nitrogen substitution
can result in low lying dark states, which can act as good donor in a photocell composite.  Recent model study has shown that control over the de-localized
 dark states can  enhance photo-currents and maximize power outputs by 35 percent over a classical cell \cite{22prl,engel}.
 
Our studies predict the emergence of high oscillator strength governed by the partial-dipole partial dipole interaction and 
 provide a comprehensive understanding and suggest guidelines for experimentalist to dope a 
graphene nano fragments  chemically to get desired properties in a controlled manner. 

We have done many body calculations considering extended Hubbard (or Pariser-Parr-Pople (PPP))  Hamiltonian as given by

\begin{eqnarray}
H= \sum_{i}n_{i}\epsilon_{i} + \sum_{i,j,\sigma} t_{i,j} (\hat{a}^{\dagger}_{i,\sigma} \hat{a}_{i,\sigma}+H.C)  \\\nonumber 
 +\frac{1}{2}\sum_{i}U_{i}n_{i \uparrow}n_{i \downarrow} +  \\\nonumber
\sum_{i,j}(V^{ppp}_{i,j}+V^{dipdip}_{i,j})(n_{i}-z_{i})(n_{j}-z_{j}) 
\end{eqnarray}


$V^{ppp}_{i,j}=14.397[(\frac{28.794}{U_i + U_j})^2+r_{ij}^2]^{-\frac{1}{2}}$

$V^{dipdip}_{i,j}=\frac{(r_{i}).(r_{j})}{r_{ij}^{3}}$

 (using standard notation) with configuration interaction (CI) method.
 
\begin{figure}[h]
\rotatebox{0}{\includegraphics*[width=14cm,height=16cm,keepaspectratio]{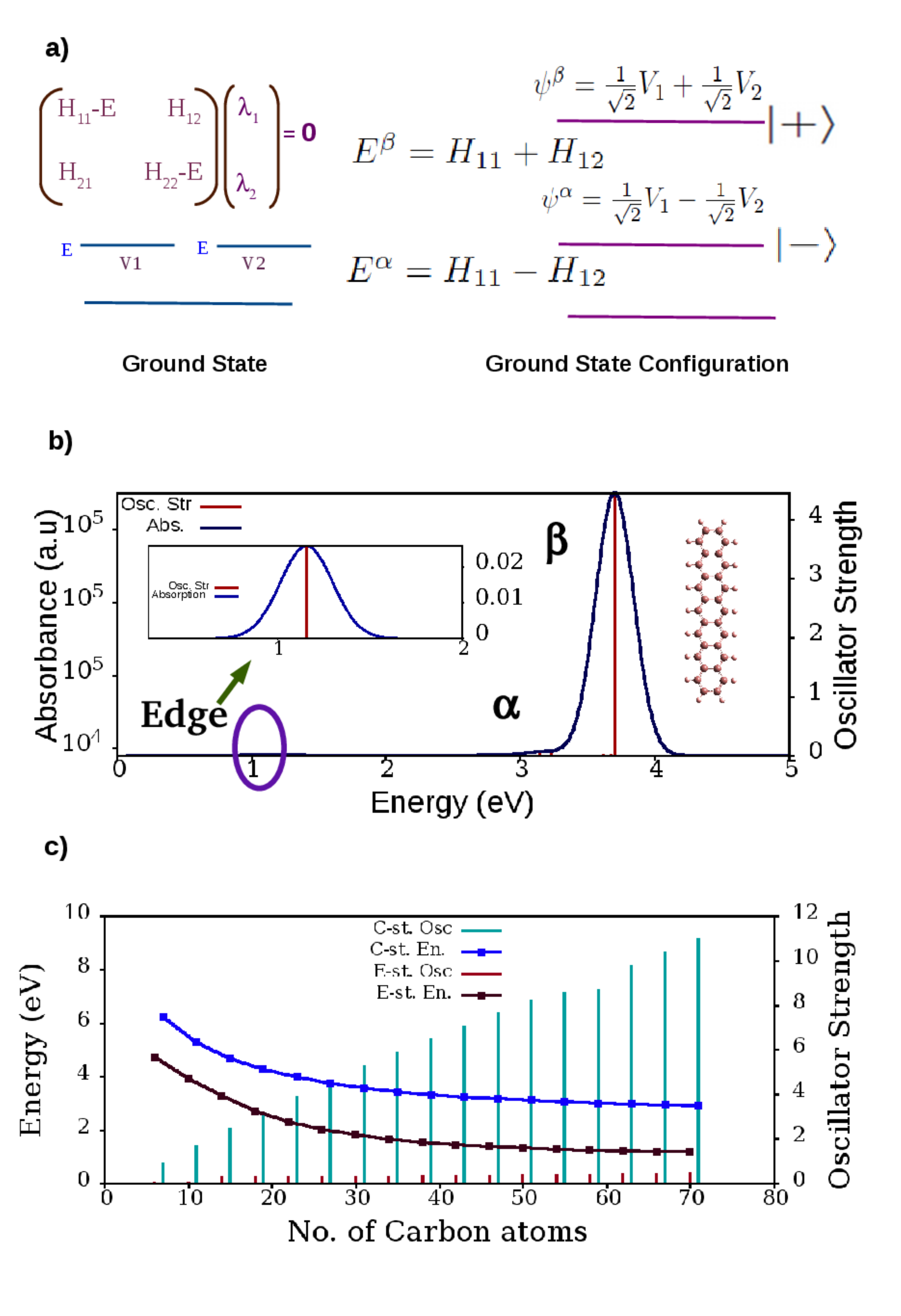}}
\caption{a) Schematic of constructive and destructive interfering states.
b)Absorption spectra of C30  acene GQD calculated by TD-DFT using B3LYP functional.
Inset shows edge states absorption.
c)With increase of size edge state absorption energy gets red shifted more steadily compared to 
coherent state $(\beta) $ absorption. For edge states 
the oscillator strengths remain more or less same but for the coherent states increase steadily (calculated from ZINDO calculation).}
\label{fig1}

\end{figure}
\section{MODELING AND COMPUTATIONAL DETAILS}
	We have considered the nearest neighbor hopping integral $t_{ij}$  to be same ($t=1 eV$) for all bonds.
	 The on site Coulomb repulsion, U  has been approximated from the difference between the first ionization energy  and the 
electron affinity of an atom, and scaled for carbon, boron and nitrogen to be  5, 4 and 7, in units of $t$ respectively. On-site energy $(\epsilon_{i})$ for carbon
nitrogen and boron are scaled as 0, -0.3 and 0.2, in units of $t$ respectively.
Since the GQDs we have considered are quite large for exact calculations, we have used CI method, which is very good for low energy optical spectra.
We have considered a small energy window in the tight binding energy spectrum and built up all possible configuration 
of 4900 or more states within the active space close to zero tight binding energy.

 Long range  Coulomb interaction potential, $V_{ij}^{ppp}$, is parametrized in Ohno interpolation scheme \cite{ohono23} and dipolar coupling,
 $V_{ij}^{dipdip} $ has been calculated from dipolar interactions \cite{ayan,pra2012dpdp}.
The dipolar coupling is dictated by the direction and the angles between the partial dipoles in GQD created by Boron or Nitrogen substitution.
This Boron or Nitrogen doping will modulate the average occupancy calculated by the number operator, resulting either in attractive potential or in repulsive potential.

\begin{table}
 \begin{tabular}{||c| c| c| c||} 
 \hline
Energy (eV) & States involved & Tr.dipole & Osc. str. \\ [0.5ex] 
 \hline\hline
3.7 (coherent ($\beta$)) & 96 -100 (43)  & 0.46461 $\uparrow$  & 4.4482 \\ 

           &  99 -103 (57) & 0.53540 $\uparrow$   &    \\
 \hline
 2.9  (incoherent ($\alpha$))& 96 -100( 56) & 0.53046 $\uparrow$ & 0.0192  \\

              & 99 -103 (42) & -0.45753 $\downarrow$ &  \\
 \hline
 1.2  (Edge) & 99 -100 (100) & 0.72121 $\uparrow$ & 0.0259 \\ [1ex] 
 \hline
\end{tabular}
\caption{For C30 acene GQD the transition energy, the states involved
for the transitions with their percentages of contribution in parenthesis,
transition dipoles with their directions, and the resulted oscillator strengths are presented }
\end{table}
We also have performed the spin-polarized density functional theory (DFT) calculations for the systems using the ab initio
 software Gaussian 09 program.  Both B3LYP and long range corrected (CAM-B3LYP) exchange correlation functional
 (which takes into account non-local correlation also) has 
 been used in association with 6-31+g(d) basis set for all
 the calculations. For large structures also Gaussian ZINDO calculations has been done.
 We have calculated the linear electronic absorption spectra using time-dependent density functional
 theory (TD-DFT) methods.  A minimum of 10 lowest excited states have been considered in all the studies. 
 
\begin{figure}[h]
\rotatebox{0}{\includegraphics*[width=15cm,height=15.8cm,keepaspectratio]{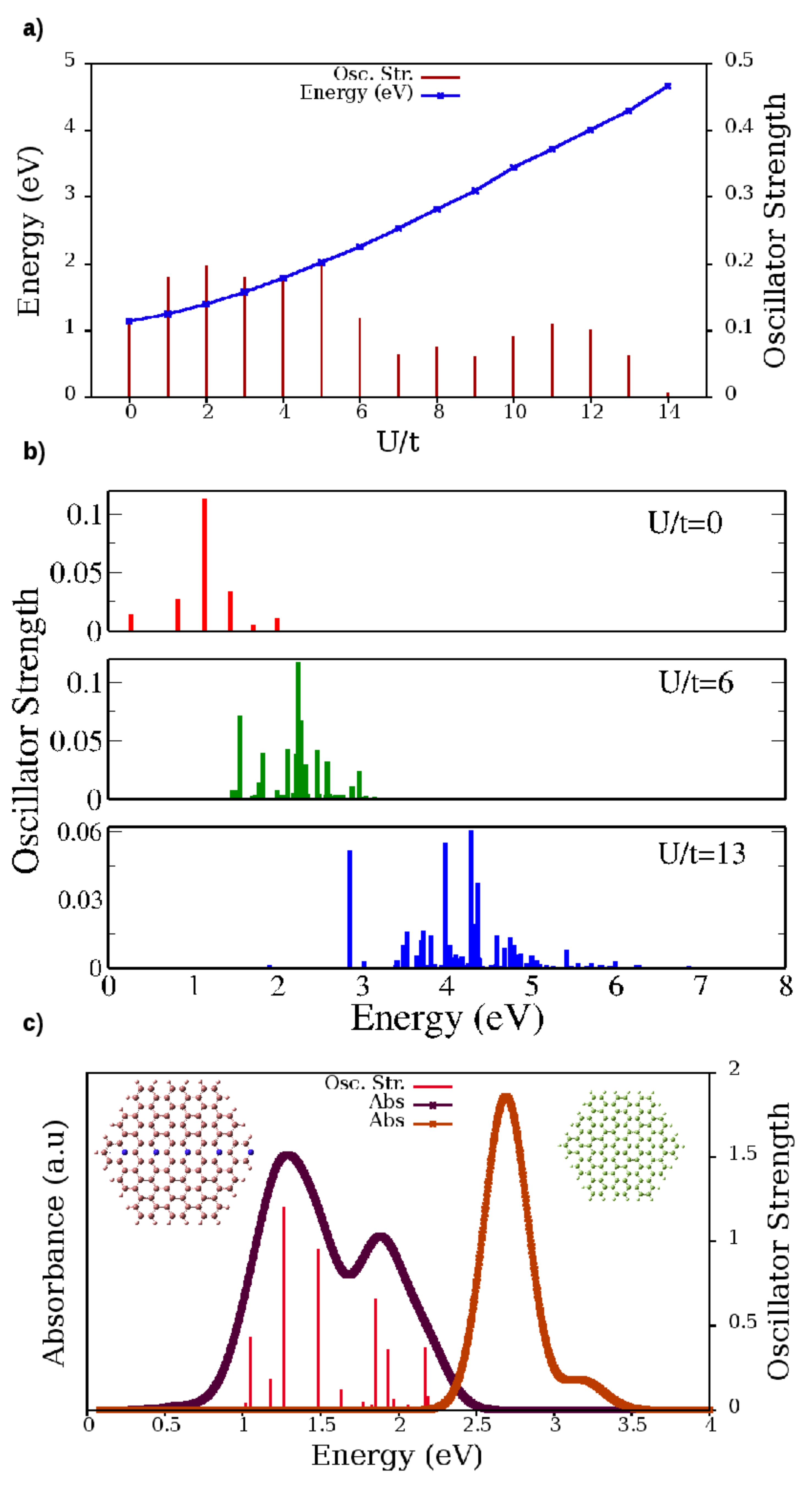}}
\caption{a) Variation of oscillator strength $(\times 10)$ and absorption energy of acene GQD with C30 at different U/t values in many body calculation. 
b) Energy states get more blue shifted, closely spaced and greater no. of states having moderate to high oscillator
strength $(\times 10)$ at high U/t ratio, where in low U/t bright states are red shifted and  more separated. c) Emergence of high oscillator strength 
at near-IR region due to partial dipole-partial dipole coupling in substituted GQD}
\label{fig2}

\end{figure}
 
\section{RESULTS AND DISCUSSION}

The key role for the high oscillator strength in the graphene nano flakes are dictated by the linear
superposition of different configurations in such a manner where transition dipoles become macroscopic due to 
constructive interference of the phase factors of  the quantum states \cite{shukla,jpc_c}. 
On the other hand, corresponding destructive interference results in the reduction of the transition dipoles producing dark states as shown in Fig.1(a).
Fig.1(b) shows the positively interfering  
coherent peak ($\beta $), negatively interfering destructive peak ($\alpha $) and edge state contributing 
peaks of 7 ring containing acene GQD with 30 carbon atoms. Table I reflects 
the transition dipole as coherent sum for the high energy peak ($\beta $) where as for 
the low energy peak ($\alpha $) the transition dipole almost cancelled out. The interesting point to note that the lowest energy peak, (shown magnified in inset) is not like additive
type rather has contribution from the low lying states, often termed as edge state transition, 
which is highly red shifted making the system near IR active. However, the oscillator strength for all the transitions and its mere presence and the strength
of edge state transition is often over estimated in DFT and ZINDO calculation (see the Supporting Information (S.I.)). 
The edge states effects are less prominent in arm chair analogue and in circularly shaped GQD. In our earlier work, we studied in details the edge state transition and 
how ground state spin multiplicity plays a crucial roles \cite{sharma}. Main point is that for all shaped GQDs, with different edges
the strength of the coherent peak increases with the system size, with lowering of transition energy (also see S.I.).

\begin{figure}[h]
\rotatebox{0}{\includegraphics*[width=20cm,height=18cm,keepaspectratio]{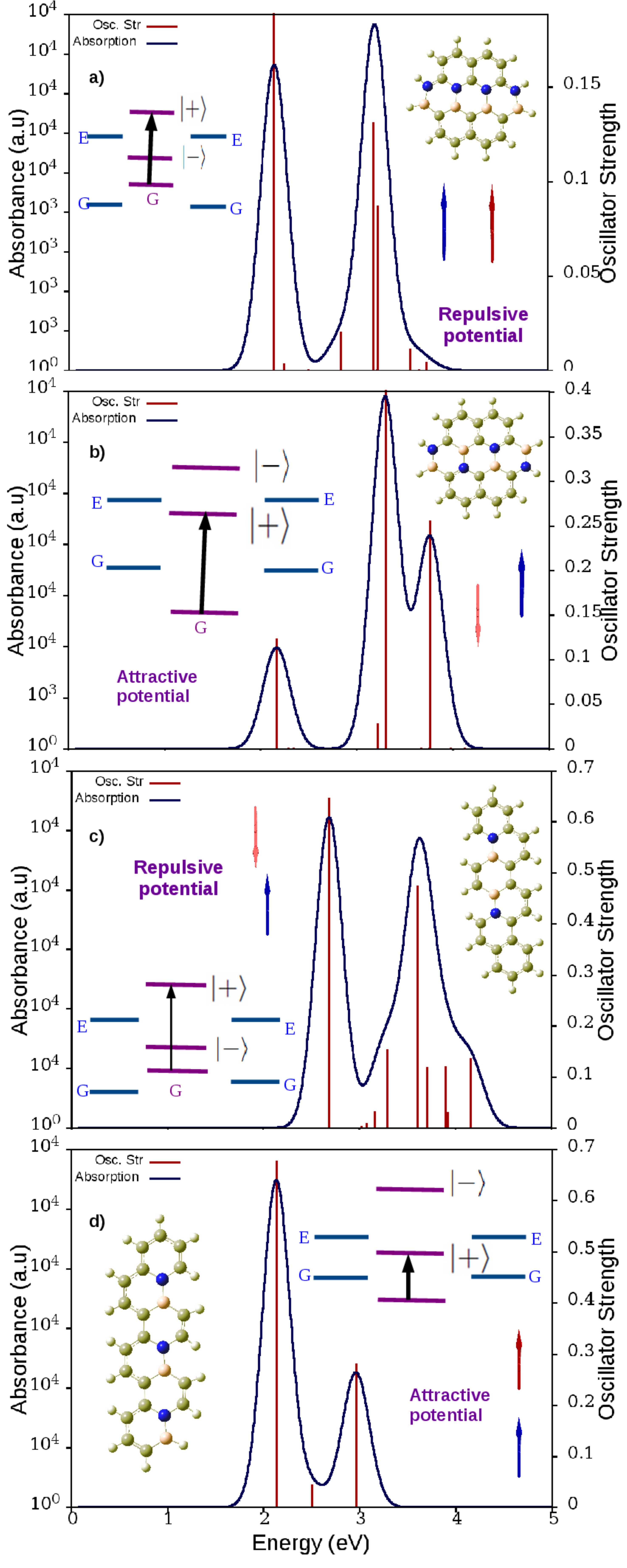}}
\caption{The orientation of boron and nitrogen in GQD dictate the alignments of partial dipoles 
to result either attractive or repulsive potential of dipolar coupling. Our DFT calculations show that for attractive partial dipole-partial dipole potential
the coherent states are more red shifted than the incoherent states, whereas for repulsive potential the incoherent states are in red end.}
\label{fig3}
\end{figure}

 Fig.1(b) shows the difference between fundamental nature of the coherent
peak and the peak generated from the edge states transition as the oscillator strengths of the edge states
remain more or less unchanged with increase of the size of the system, while the oscillator strengths of
the coherent peaks increase steadily with the system size due to more and more positive in-phase combination of 
the transition dipoles of the de-localized states (also see S.I.). 
 
We believe introduction of strain in graphene quantum dots can alter the on site Coulombic interaction. The effect of strain can be 
utilized to modify the energy and the transition strength of an optical probe.
Fig.2(a) shows how transition energy and the transition amplitude depends on the U/t.
With the increase in strain ie the higher U/t, one can achieve high energy transition but at the expense of transition strength.
Fig.2(b) shows that at low  $U/t$, the whole spectrum is red shifted with fewer numbers of bright states, which are widely separated. 
On the other hand,  at high $U/t$ limit, the whole spectrum is blue shifted with more number of closely spaced bright states  and many transitions of moderate to good 
oscillator strengths. At high $U/t$ interaction, degeneracy of some dark states lift, resulting in some states with favoured parity for optical transition.
This kind of closely spaced bright states can be interesting for coherent excitation.

\begin{table}
 \begin{tabular}{ || m{8em} | m{2cm}| m{2cm} | m{1cm} ||}
 \hline
Energy (eV) & States involved & Tr.dipole & Osc. str. \\ [0.5ex] 
 \hline\hline
3.18 (blue shifted coherent($\beta$))
& 76 - 79 (57)  & 0.53422 $\uparrow$  & 0.1308 \\ 
  &  78 - 81 (29)  & 0.38223 $\uparrow$   &    \\
     \hline
 2.84(red shifted incoherent($\alpha$))
  &  78 - 81 (62) & 0.55534 $\uparrow$ &  0.02 \\
 &  76 - 79  (32)  &-0.40110 $\downarrow$ &  \\
 \hline
2.14 (edges state)               
 &  78 - 79(96) & 0.69149 $\uparrow$ &  0.1885 \\ 
[1ex] 
 \hline
\end{tabular}
\caption{For GQD shown in Fig 3a, $i.e.$ chemical doped GQD with repulsive potential, the transition energy, the states involved
for the transitions with their percentages of contribution in parenthesis,
transition dipoles with their directions, and the resulted oscillator strengths are presented}
\end{table}

\begin{table}
 \begin{tabular}{ || m{8em} | m{2cm}| m{2cm} | m{1cm} ||}
 \hline
Energy (eV) & States involved & Tr.dipole & Osc. str. \\ [0.5ex] 
 \hline\hline
2.96(red shifted coherent($\beta$))
& 85 - 87 (50)  & 0.49758 $\uparrow$  & 0.2796 \\ 
  &  86 - 89 (48)  & 0.49243  $\uparrow$   &    \\
     \hline
 3,27 (blue shifted incoherent($\alpha$))
  &  85 - 87 (44) & 0.46823 $\uparrow$ & 0.0203 \\
 & 86 - 89  (42)  & -0.46064$\downarrow$ &  \\
 \hline
 2.13 (edges state)               
 & 86-87 (100) & 0.70557 $\uparrow$ & 0.6766 \\ 
[1ex] 
 \hline
\end{tabular}
\caption{For GQD shown in Fig 3d, $i.e.$ chemical doped GQD with attractive potential, the transition energy, the states involved
for the transitions with their percentages of contribution in parenthesis,
transition dipoles with their directions, and the resulted oscillator strengths are presented}
\end{table}

We have searched how the delocalized states and the transition strength can be tuned through the implementation of chemical doping of nitrogen and/or boron in GQD. 
The key role is played by the partial dipole -partial dipole coupling.

\begin{figure}[h]
\rotatebox{0}{\includegraphics*[width=20cm,height=18cm,keepaspectratio]{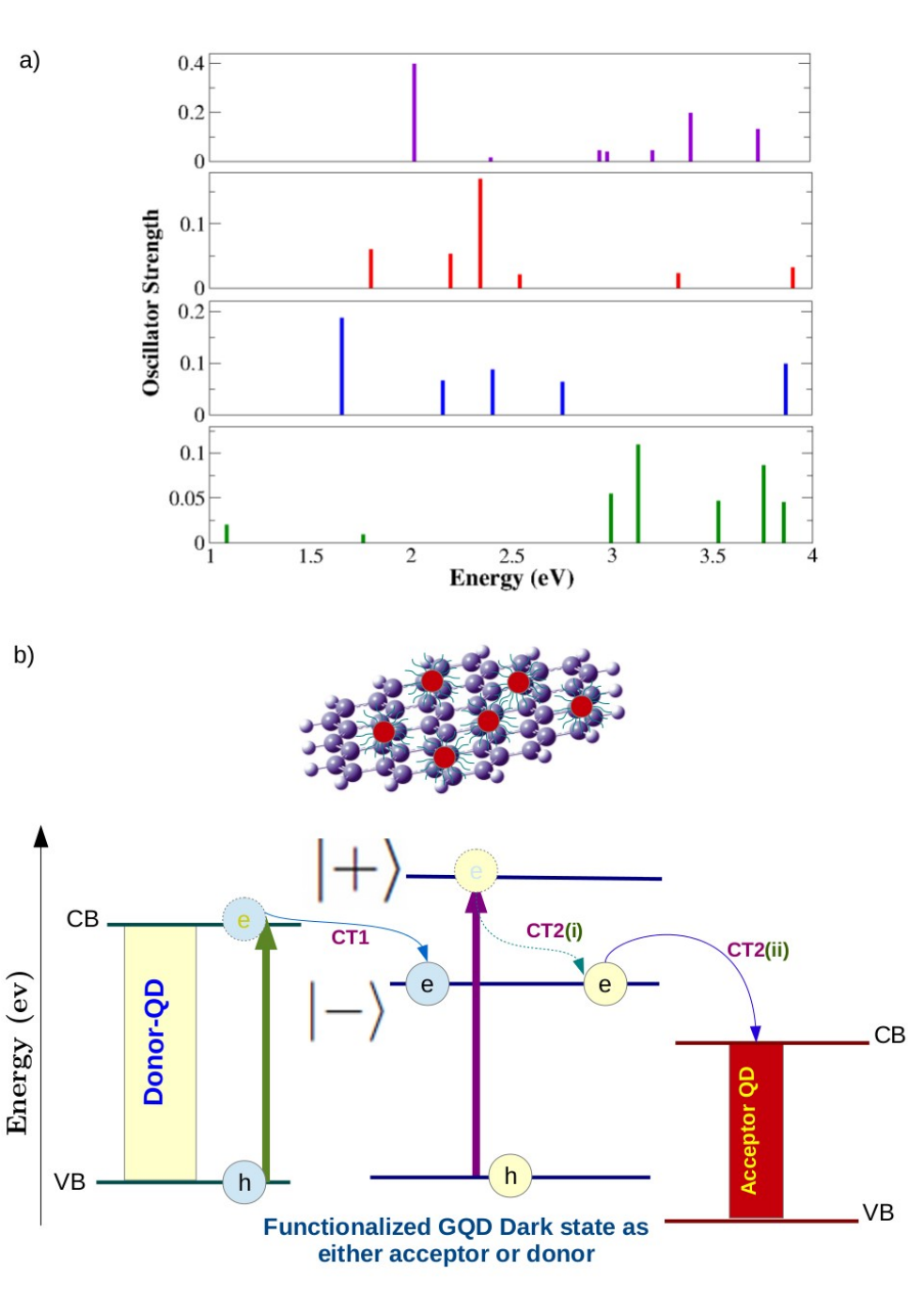}}
\caption{a)For the same sequence of GQD shown in Fig.3, the optical transition calculated by many body calculation
depends in similar fashion with DFT. Emergence of high oscillator strength $(\times 10)$ in the lower energy is observed due to 
dipole-dipole interaction. b)Schematic of the hybrid QDS and energy states involved for electron transfer.
The relative position of dark state in lower energy compare to the bright state in GQD 
can make the composite efficient in electron transfer either by taking electron from other excited 
QD (CT1) or for GQD excitation, electron from bright state can go through phonon relaxation to the dark state
prohibiting the radiative loss and enhancing the electron transfer.}
\label{fig4}
\end{figure}

 Fig.2(c) and Fig.3 show how the different kind of orientation of N, or B and N substitution can control the alignment of 
partial dipole to produce either attractive or repulsive potential of dipolar coupling and can result in high oscillator strength at new energy.
Fig.3 shows the structures of GQDs along with their partial dipole alignment, and the transition energy and strength obtained from DFT calculations.
For the same structures, the C.I calculations are shown in Fig.4(a) in the same sequence.
For some other structures and larger GQDs, the results of many body, DFT and semi-empirical calculations are shown in S.I.
Fig.3 in-sets show the schematic of two states model for dipolar coupling  $i.e.$ Davydov  splittings, where the relative position of coherent and incoherent states are obtained from DFT calculations,
presented in Table 2 and 3 (also See S.I. for other structures). 
  Davydov splitting is mainly discussed in the aggregate
type of systems\cite{davy_extra}. Where the quasi-classical vector treatment
and the electrostatic interaction between the transition moment dictates the splitting energy between the coherent and the incoherent states ( See S.I.).
Interestingly, here also, we find similar kind of splitting due to the partial dipole-partial dipole coupling in doped GQD systems. Our DFT calculations 
on GQDS gives the relative position of coherent and incoherent states in same way as in the quasi-classical vector model treatment for aggregates. 
Table 2 and 3 reflects the states involved and their contribution in transition dipoles for either attractive or repulsive potentials for two GQDS, for more GQDS see S.I.
 For the attractive dipolar coupling the constructive interfering states get shifted to lower energy compared to destructive interfering states.
In case of repulsive dipolar coupling  the dark states get red shifted. Interesting point to note that in case of substitution here the edge states contributions are also enhanced.
 The many body calculation with extended Hubbard model 
Hamiltonian shows qualitatively the similar effects as observed for DFT calculations. The emergence of high oscillator strength at the red end is also observed here
depending on the attractive dipole dipole interactions. Thus, in substituted GQD, for attractive potential, the bright states are more red shifted than the dark states, 
where as for repulsive potential, dark states are generated at the red end.

Interestingly, experimentalists are using group IV -VI quantum dots along with graphene matrix or graphene quantum dots to prepare hybrid
quantum dots nano films where one acts as donor and the other as acceptor and looking for their electron transfer dynamics \cite{kamat,tonu2015}. Efficient
electron transfer can dramatically increase photo current generation in the hybrid QD sensitized solar cells. In fact, when graphene quantum 
dot is photo-excited, the exciton generated  need to migrate to donor acceptor interface to dissociate.

$D^*+A\rightarrow D+A^*$

$k_{exchange}\propto {\arrowvert\langle\psi^*(D)  \psi(A)  \arrowvert H_{ex} \arrowvert\psi(D)  \psi^*(A)  \rangle \arrowvert}^2$

$H_{ex} \propto e^{-R_{DA}}$

$k_{dipdip}\propto {\arrowvert\langle\psi^*(D)  \psi(A)  \arrowvert H_{dipdip} \arrowvert\psi(D)  \psi^*(A)  \rangle\arrowvert}^2$

$H_{dipdip}\propto\frac{\mu^*_D\mu^*_A}{R^3_{DA}}$

Energy transfer can occur through either dipole-dipole interaction or electron-exchange interaction. Energy transfer through 
electron exchange interaction (require orbital overlap, Dexter mechanism) and energy transfer through dipole-dipole interaction 
(Froster mechanism) are shown in above equation\cite{Fortser, Dexter,scholes}. In exchange mechanism, the rate of energy transfer decreases exponentially with the increase in donor acceptor distance (0.1-1 nm) whereas 
through the dipole-dipole interaction, energy transfer can occur over long range (1-10 nm).
 
Now, as we have shown, engineering the GQDs by site specific chemical doping, one can introduce the dipolar interaction in controlled
 manner to get red end dark states important for efficient electron transfer and that can be used for GQD sensitized solar cell.  In Fig.4(b) we present the schematic of 
electron transfer and charge separation in GQD-QD composite and show how red end incoherent state facilate electron transfer. 
Here, either in CT1 path way, the dark state of doped GQD can accept the electron from a photo excited donor QD, 
or in CT2 pathway, photo excited GQD can relax to the dark state followed by efficient electron transfer to other acceptor QD, resulting in efficient charge separation.
Again, also the bright state can be tuned to high oscillator strength for  enhanced photonic activity.

\section{CONCLUSIONS}	Our studies establish that controlling the coherent superposition of quantum states,  the emergence of high oscillator
	strength in desirable energy can be achieved  through partial dipole-partial dipole interaction in substituted graphene quantum dots.
	Our DFT and many body calculations show that in substituted GQDs due to partial-dipole partial dipole coupling, Davydov type of splittings are obtained
	in same fashion as observed with the quasi classical vector treatment of dipolar coupling in aggregates.
	We have shown that in attractive dipolar potential, the bright state in functionalized GQD is more red shifted compared to its dark counter part.
		Whereas the repulsive dipolar potential produces red end dark state which can be used for efficient electron transfer, 
	as has been reported earlier that \cite{22prl,engel}, a red shifted dark state compared to its bright counter part can result in enhanced photocell efficiency.
		Experimentalists have reported efficient electron transfer in the low lying dark state of nano structured graphene from group VI-VII quantum dots \cite{kamat,tonu2015}.
	Hence, utilizing partial dipole-partial dipole coupling, proper architecture of the dark state in GQD, relative to the other QDs in hybrid QD system can be achieved, 
	paving the path for efficient electron transfer.  
	Again our results reflects that the non additive edge states transition can be enhanced through functionalization of GQDs.
	Also, our studies reveal that the effect of strain can increase the on site repulsion to get more closely spaced bright states which are interesting for coherent excitation.

\section{Supporting Information} Additional computational results with comparison of different levels of theory, characteristics of coherent transition of different shaped GQDS from DFT, 
Effects of strain on optical transition by many body calculation for other structures, dependence of transition energy and strength on the system shape and sizes, the details of transition energies and the states 
involved for the transitions and the nature of Davydov splitting for all the structures considered,  some more examples of orientation of partial dipoles in substituted GQDs
to control the relative position of coherent and incoherent states are presented in the supporting information.
This material is available free of charge via the Internet at http://pubs.acs.org.

\section{ACKNOWLEDGMENTS} We thank S. S. R. K. C. Yamijala and S. Banerjee for discussions. S.K.P. acknowledges DST for funding. B.P. acknowledges UGC for fellowship.

\end{document}